\documentclass[twocolumn,prl,showpacs,superscriptaddress,preprintnumbers,amsmath,amssymb,floatfix]{revtex4}
\usepackage{graphicx}
\usepackage{dcolumn}

\usepackage{bm}

\begin{document}

\title{The $c$-axis dimer and its electronic break-up: the insulator-to-metal transition in Ti$_2$O$_3$}

\author{C.~F.~Chang}
      \affiliation{Max Planck Institute for Chemical Physics of Solids, N{\"o}thnitzer Stra{\ss}e 40, 01187 Dresden, Germany}
\author{T.~C.~Koethe }
     \affiliation{Institute of Physics II, University of Cologne, Z{\"u}lpicher Stra{\ss}e 77, 50937 Cologne, Germany}
\author{Z.~Hu}
     \affiliation{Max Planck Institute for Chemical Physics of Solids, N{\"o}thnitzer Stra{\ss}e 40, 01187 Dresden, Germany}
\author{J.~Weinen}
     \affiliation{Max Planck Institute for Chemical Physics of Solids, N{\"o}thnitzer Stra{\ss}e 40, 01187 Dresden, Germany}
\author{S.~Agrestini}
     \affiliation{Max Planck Institute for Chemical Physics of Solids, N{\"o}thnitzer Stra{\ss}e 40, 01187 Dresden, Germany}
\author{J.~Gegner}
     \affiliation{Institute of Physics II, University of Cologne, Z{\"u}lpicher Stra{\ss}e 77, 50937 Cologne, Germany}
\author{H.~Ott}
     \affiliation{Institute of Physics II, University of Cologne, Z{\"u}lpicher Stra{\ss}e 77, 50937 Cologne, Germany}
\author{G.~Panaccione}
       \affiliation{TASC Laboratory, IOM-CNR, Area Science Park, S.S.14, Km 163.5, I-34149 Trieste, Italy}
\author{Hua~Wu}
      \affiliation{Laboratory for Computational Physical Sciences (MOE), State Key Laboratory of Surface Physics,
                     and Department of Physics, Fudan University, Shanghai 200433, P.R. China}
\author{M.~W.~Haverkort}
      \affiliation{Institute for Theoretical Physics, Heidelberg University, Philosophenweg 19, 69120 Heidelberg, Germany}
\author{H.~Roth}
      \affiliation{Institute of Physics II, University of Cologne, Z{\"u}lpicher Stra{\ss}e 77, 50937 Cologne, Germany}
\author{A.~C.~Komarek}
       \affiliation{Max Planck Institute for Chemical Physics of Solids, N{\"o}thnitzer Stra{\ss}e 40, 01187 Dresden, Germany}
\author{F.~Offi}
       \affiliation{CNISM and Dipartimento di Scienze, Universit\`{a} Roma Tre, Via della Vasca Navale 84, I-00146 Rome, Italy}
\author{G.~Monaco$^{\ast}$}
       \affiliation{European Synchrotron Radiation Facility, BP220, 38043 Grenoble, France}
\author{Y.-F.~Liao}
     \affiliation{National Synchrotron Radiation Research Center, 101 Hsin-Ann Road, Hsinchu 30076, Taiwan}
\author{K.-D.~Tsuei}
     \affiliation{National Synchrotron Radiation Research Center, 101 Hsin-Ann Road, Hsinchu 30076, Taiwan}
\author{H.-J.~Lin}
     \affiliation{National Synchrotron Radiation Research Center, 101 Hsin-Ann Road, Hsinchu 30076, Taiwan}
\author{C.~T.~Chen}
     \affiliation{National Synchrotron Radiation Research Center, 101 Hsin-Ann Road, Hsinchu 30076, Taiwan}
\author{A.~Tanaka}
      \affiliation{Department of Quantum Matter, ADSM, Hiroshima University, Higashi-Hiroshima 739-8530, Japan}
\author{L.~H.~Tjeng}
	\affiliation{Max Planck Institute for Chemical Physics of Solids, N{\"o}thnitzer Stra{\ss}e 40, 01187 Dresden, Germany}

\date{\today}

\begin{abstract}

We report on our investigation of the electronic structure of Ti$_2$O$_3$ using (hard) x-ray photoelectron and
soft x-ray absorption spectroscopy.  From the distinct satellite structures in the spectra we have been able to
establish unambiguously that the Ti-Ti $c$-axis dimer in the corundum crystal structure is electronically present
and forms an $a_{1g}a_{1g}$ molecular singlet in the low temperature insulating phase. Upon heating we
observed a considerable spectral weight transfer to lower energies with orbital reconstruction. The insulator-metal
transition may be viewed as a transition from a solid of isolated Ti-Ti molecules into a solid of electronically partially
broken dimers where the Ti ions acquire additional hopping in the $a$-$b$ plane via the $e_g^{\pi}$ channel,
the opening of which requires the consideration of the multiplet structure of the on-site Coulomb interaction.

\end{abstract}

\pacs{}

\maketitle


The role of ion pair formation for the metal-insulator transition (MIT) in early transition metal oxides with the
octahedra sharing either a common face or a common edge has been a matter of debate in the past several
decades \cite{Goodenough1960,Zeiger-PRB-1975,Zandt1968,Castellani1978,Rice1995,Mattheiss1996,Rice2000,Park2000,Rice2001,
Elfimov2003,Tanaka2004,Poteryaev2004,Eyert2005,biermann2005,Haverkort2005,Koethe2006,Hitoshi2006,Uchida2008}.
Based on the presence of the $c$-axis V--V dimers in the corundum crystal structure of V$_2$O$_3$,
C. Castellani \textit{et al.} \cite{Castellani1978} proposed a molecular singlet model for the $a_{1g}$ orbitals,
projecting the system effectively onto a solid with $S=1/2$ entities which then should carry the essential
physics for the MIT and the magnetic structure in the antiferromagnetic insulating phase. However, soft x-ray
absorption spectroscopy (XAS) experiments \cite{Park2000} showed that the two $d$
electrons on each V are in the high-spin $S=1$ state, implying that the atomic Hund's rule coupling is much
stronger than the intra-dimer hopping integrals.  Furthermore, using band structure calculations Elfimov
\textit{et al.} \cite{Elfimov2003} found that the intra-dimer hopping integral is not the most important one,
rather that the hopping integrals between second, third and fourth nearest V neighbors are at least equally
important: in other words, the $c$-axis dimers need not to be present electronically although structurally they
are there.

\begin{figure}
    \centering
    \includegraphics[width=8.5cm]{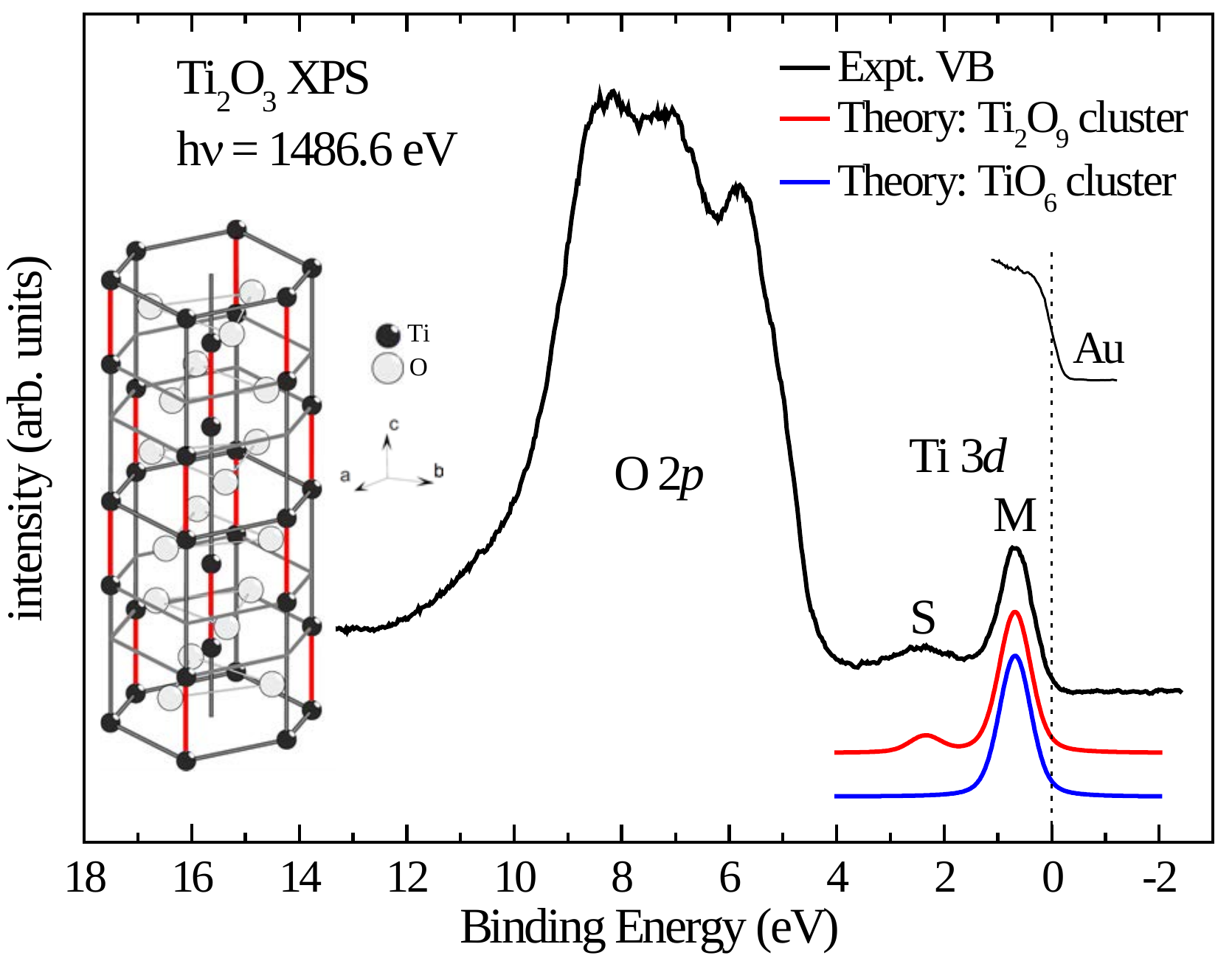}
    \caption{Experimental valence band (VB) XPS spectrum of Ti$_2$O$_3$ taken at 300 K (black line),
     together with the Au Fermi cut-off (thin black line) as reference for $E_F$. Also shown is the
     Ti $3d$ one-electron removal spectrum from a TiO$_6$ (blue line) and Ti$_2$O$_9$ (red line)
     cluster as described in the text. Inset: Corundum structure of Ti$_2$O$_3$ with the
     $c$-axis Ti-Ti dimers marked in red.}
    \label{Fig1_VB_CI-dimer}
\end{figure}

Ti$_2$O$_3$ shares much of the fascination as V$_2$O$_3$. It has also the corundum crystal structure
(see the inset in Fig.~\ref{Fig1_VB_CI-dimer}) and exhibits upon lowering the temperature a MIT \cite{Morin1959}.
The earliest models explained the low temperature insulating phase of Ti$_2$O$_3$ by assuming a band
splitting caused by an antiferromagnetic long-range order \cite{Morin1959}. However, different from V$_2$O$_3$,
the transition is gradual and is not accompanied by a structural transition nor magnetic ordering
\cite{Keys1967,Moon1969,Robinson1977}. It was also proposed by Goodenough and van-Zandt \textit{et al.} that the short $c$-axis pair bond length, which is
with 2.578 {\AA} at 300 K \cite{Abrahams1963} much shorter than in V$_2$O$_3$ with 2.697 {\AA} at 300 K
\cite{Dernier1970}, increases the trigonal crystal field splitting so that the conductivity gap is opened \cite{Goodenough1960,Zandt1968}.
This model is, however, contradicted by band structure calculations which showed that the overlap of the $a_{1g}$
and $e_{g}^{\pi}$ orbitals can only be suppressed for an unrealistically short bond length \cite{Mattheiss1996},
i.e. Ti$_2$O$_3$ is a metal at all temperatures with mixed $a_{1g}$-$e_{g}^{\pi}$ states for the $c$-axis dimer
from the point of view of band theory \cite{Zeiger-PRB-1975,Mattheiss1996,Eyert2005}. Correlation effects
have to be included in one way or another to explain the insulating ground state in Ti$_2$O$_3$
\cite{Tanaka2004,Poteryaev2004,Iori-PRB-2012,Guo-JPhysCondMat-2012,Singh_JAlloysCompounds-2016}.

Here we report on our spectroscopic study of the electronic structure of Ti$_2$O$_3$ with the goal to determine
whether and how correlation effects and the $c$-axis dimer play a role for the formation of the low-temperature
insulating phase. Moreover, we would like to identify the key factors in the electronic structure that can transform
the compound from an insulator into a metal. We found in our spectra direct evidence that Ti$_2$O$_3$ is a
strongly correlated system in which the $c$-axis Ti-Ti dimers form isolated $a_{1g}a_{1g}$ molecular singlets at
low temperatures, and that at high temperatures the dimers partially break-up electronically with the Ti ions gaining
hopping in the $a$-$b$ plane via the $e_g^{\pi}$ channel. Crucial is that the orbital switching from $a_{1g}$
towards $e_g^{\pi}$ is possible only if the multiplet aspect of the on-site Coulomb interaction is another decisive
element in the electronic structure of Ti$_2$O$_3$.


X-ray photoelectron spectroscopy (XPS) measurements with $h\nu=1486.6$ eV were performed in Cologne
using a Vacuum Generators twin crystal monochromatized Al-$K\alpha$ source and a Scienta SES-100 electron
energy analyzer. The overall energy resolution was set to 0.4 eV. Hard x-ray photoelectron spectroscopy (HAXPES)
experiments were carried out at the ID16 beamline of the ESRF using the VOLPE spectrometer with
$h\nu= 5931$ eV and an overall resolution of 0.4 eV, as well as at the Taiwan beamline BL12XU at SPring-8
using the Max-Planck-NSRRC end-station equipped with a MB Scientific A-1 HE hemispherical analyzer
employing $h\nu\simeq 6.5$ keV and an overall resolution of 0.2 eV. Soft-x-ray absorption spectra (XAS) were
collected at the Dragon beam line at the NSRRC in Taiwan in the total electron yield mode with a photon
energy resolution of 0.25 eV and degree of linear polarization of ~98\%. All spectra were collected from
freshly \emph{in vacuo} cleaved Ti$_2$O$_3$ single crystals. Ti$_2$O$_3$ single crystals were grown by using
the floating zone method. The purity and structure of crystals were verified as a single phase crystal by using
EDX, powder diffraction measurements, Laue and polarization microscopy. The stoichiometry of the crystals
has been characterized by thermogravimetric analysis.


Figure~\ref{Fig1_VB_CI-dimer} shows the valence band XPS spectrum of Ti$_2$O$_3$ taken at 300 K,
i.e. in the insulating phase. The group of peaks at 4--11 eV binding energies is mainly the contribution
of O $2p$ states. The lower binding energy region from the Fermi level up to 4 eV consists of mainly the Ti~$3d$.
This part of the spectrum is characterized by two distinct spectral features. The main line (M) is a quite
symmetric peak centered at about 0.68 eV with a width of $\approx 0.8$ eV (FWHM). The semiconducting
or insulating nature of Ti$_2$O$_3$ at room temperature \cite{Zandt1968,Honig68a,Honig68b} is reflected
by the fact that the spectral weight vanishes at the Fermi level, in agreement with earlier photoemission
reports \cite{Kurtz1982,Mckay1987,Smith1988,Smith88b,Uozumi93,Kotani1996} and the observation of
a 0.2 eV gap in the optical conductivity \cite{Lucovsky1979,Uchida2008}.

The second feature is a somewhat broader but clearly noticeable satellite (S) at around 2.43 eV binding energy.
The origin of this peak has so far been disputed. Ultraviolet photoelectron spectroscopy (UPS) studies had speculated
that it was a surface state with a considerable $3d$ band character \cite{Kurtz1982,Mckay1987}. However,
an angle-resolved UPS study was not able to confirm this speculation \cite{Smith1988}. We claim here
that our spectrum is representative for the bulk material, i.e. that both features M and S belong to the
photoemission spectrum of bulk Ti$_2$O$_3$. This relies on the fact that our spectrum was taken on a cleaved
single crystal at normal emission with a photon energy of 1486.6 eV, thereby obtaining larger probing depths
\cite{Weschke1991,Sekiyama2000,Sekiyama2004}. Below we will also provide more spectroscopic evidence
that all our spectra are representative for the bulk.

The absence of any spectral weight at the Fermi level in the low temperature phase of Ti$_2$O$_3$
invalidates the predictions of band structure calculations \cite{Zeiger-PRB-1975,Mattheiss1996,Eyert2005}
which always display a finite density of states at the Fermi level. This is a strong sign that correlation effects play
a crucial role. Indeed, using hybrid functionals, non-local exchange, or dynamical mean field approaches (DMFT),
one can recover a band gap in the calculations
\cite{Poteryaev2004,Iori-PRB-2012,Guo-JPhysCondMat-2012,Singh_JAlloysCompounds-2016}.
Interestingly, all calculations except the two-site cluster DMFT \cite{Poteryaev2004} did not produce a satellite
structure like the feature S we observe at 2.43 eV binding energy. Somewhat puzzling is yet that the two-site
cluster DMFT calculation did produce a satellite structure, but with a much too low intensity.

In order to unveil the origin of the satellite structure S, we now resort to configuration interaction cluster
calculations with full atomic multiplet theory, an approach which is very successful to explain quantitatively
the basic features in many photoelectron and x-ray absorption spectra of $3d$ transition metal oxides
\cite{Fujimori-Minami-1984,Tanaka1994,DeGroot1994,Thole1997}.
We start with the standard single transition metal site cluster, i.e. a TiO$_6$ octahedral cluster with the Ti ion
in the center \cite{Kotani1996}, and use
model parameters typical for titanium oxides \cite{Tanaka1994,Bouquet-Fujimori-1992,Poteryaev2004,CIpara}.
We find that the Ti $3d$ one-electron removal spectrum near the Fermi level consists of a single peak,
see the blue curve in Fig.~\ref{Fig1_VB_CI-dimer}. Satellite S is not reproduced in a single-site cluster. Next,
motivated by the presence of the Ti-Ti $c$-axis dimer in the crystal structure, we calculate the spectrum of a
Ti$_2$O$_9$ cluster consisting of two face-shared TiO$_6$ octahedral units along the c-axis,
using the same parameters as for the TiO$_6$ cluster, but with the addition of a
parameter describing the inter-Ti hopping \cite{CIpara}. The result is given by the red curve
in Fig.~\ref{Fig1_VB_CI-dimer} and we can observe that both the satellite structure S and the main peak M
can be reproduced very well.

To interpret this result, we can use the following schematic model. The relevant orbital in this Ti-Ti
dimer is the one pointing along the bond, namely the $a_{1g}$. With each Ti ion having the 3+ valence,
we then consider the following singlet configurations forming the ground state: $a_{1gA}$$a_{1gB}$,
$a_{1gA}$$a_{1gA}$, and $a_{1gB}$$a_{1gB}$, where $A$ and $B$ denotes the two Ti sites.
The configurations with the double occupation on one site have the extra energy Hubbard $U$, and
the hopping integral between the $a_{1gA}$ and $a_{1gB}$ orbitals is denoted by $t$ \cite{Ashcroft}.
The photoemission final states, in which one electron has been removed, are given by the following two
configurations, namely $a_{1gA}$ and $a_{1gB}$, which are degenerate in energy and form bonding
and anti-bonding states with energies -$t$ and +$t$. Their energy separation, $2t$, can then be read
directly from the energy separation between feature M and S, i.e. $2t=1.75$ eV ($t=0.88$ eV).
The intensity ratio between M and S is determined by $U/t$. For $U/t=0$, the intensity of the satellite S
vanishes and we are back in a one-electron approximation. In the limit of $U/t \rightarrow \infty$,
M and S will have equal intensities. From the experimental intensity ratios, we estimate that $U/t$ is
about $3-4$, i.e. $U \approx 2.5-3.5$\,eV. The essential outcome of the Ti$_2$O$_9$ cluster calculation
is thus that the inter-site Ti hopping together with the on-site Coulomb interaction produces a main line M
with a satellite structure S, and that the presence of M and S shows that there is a strong electronic bond
between the two Ti ions of the dimer.

\begin{figure}
    \centering
    \includegraphics[width=8.5cm]{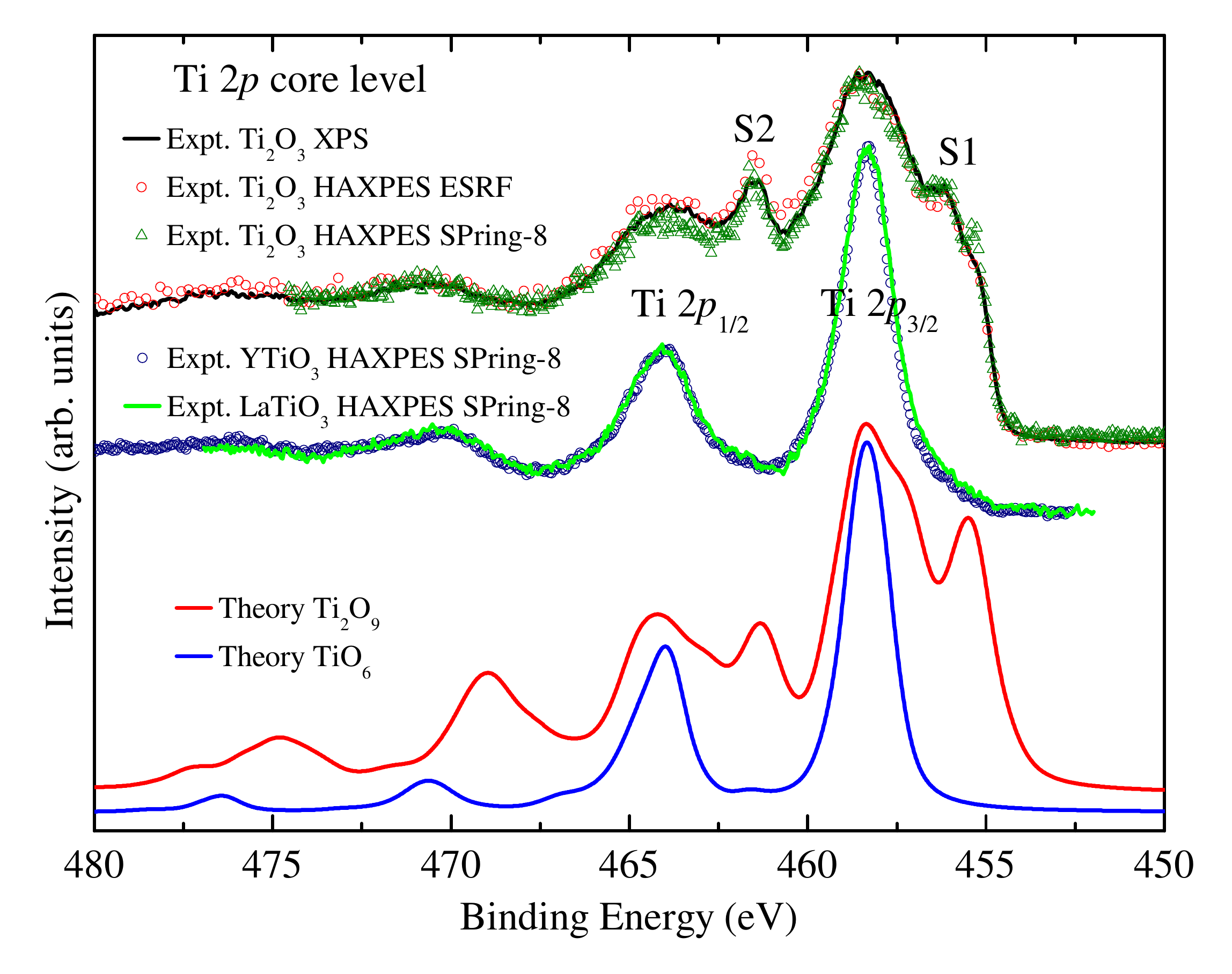}
    \caption{Experimental Ti $2p$ core-level photoemission spectra of  Ti$_2$O$_3$ taken at 300 K
    with $h\nu=1486.6$ eV (black line, XPS),
    with $h\nu= 5931$ eV (red circle, HAXPES ESRF), and
    with $h\nu\simeq 6500$ eV (dark green triangle symbol, HAXPES SPring-8), and
    experimental Ti $2p$ core-level photoemission spectra of
    YTiO$_3$ taken at 300 K (navy circle) and of LaTiO$_3$ taken at 200 K (green line)
    with $h\nu\simeq 6500$ eV (HAXPES SPring-8).
    Also shown are the theoretical configuration interaction calculations using a
    TiO$_6$ (blue line) and a Ti$_2$O$_9$ (red line) cluster, see text.}
    \label{Fig:2p}
\end{figure}

In order to collect more evidence for the presence of the strong electronic bond within the Ti-Ti $c$-axis dimer
we now investigate the Ti $2p$ core level spectrum. Fig.~\ref{Fig:2p} shows the Ti $2p$ core level
spectrum of Ti$_2$O$_3$ taken at 300 K with $h\nu=1486.6$ eV (black line, XPS), with $h\nu= 5931$ eV
(red circle, HAXPES ESRF), and with $h\nu\simeq 6500$ eV (dark green triangle symbol, HAXPES SPring-8)
together with the Ti $2p$ spectra from YTiO$_3$ (navy circle) and LaTiO$_3$ (green line)
with $h\nu\simeq 6500$ eV (HAXPES SPring-8). The first aspect we would like to mention is that the Ti$_2$O$_3$
XPS spectrum is identical to the bulk-sensitive Ti$_2$O$_3$ HAXPES spectra taken at ESRF and SPring-8.
This demonstrates that our XPS spectra, i.e. also the one displayed in Fig.~\ref{Fig1_VB_CI-dimer}, are
representative for the bulk material. The Ti $2p$ core level XPS spectra reported so far in the literature \cite{Kotani1996,Kurtz1998} have a lineshape different from ours. We have therefore carried out the Ti $2p$ experiment multiple times, using different batches of Ti$_2$O$_3$ samples, and  using the XPS in our home laboratory as well as the more bulk sensitive HAXPES at two different experimental stations (ESRF and SPring8), all to verify that the spectra we were collecting are indeed reproducible. The second aspect to notice, is that the Ti$_2$O$_3$ spectra are
very different from those of YTiO$_3$ and LaTiO$_3$, despite the fact that all are Ti$^{3+}$ $3d^1$
compounds. The satellites marked as S1 and S2 have truly massive intensities, indicative of essential
differences in the local electronic structure between Ti$_2$O$_3$ and YTiO$_3$/LaTiO$_3$.

To quantify the observations, we calculate the Ti $2p$ electron removal spectrum using the single-site
Ti cluster, i.e. TiO$_6$, and the two-site Ti cluster, i.e. Ti$_2$O$_9$, as described above.
The result for the TiO$_6$ cluster is shown by the blue line in Fig. \ref{Fig:2p}: the calculated spectrum is essentially similar
to the one reported in Ref.~[\onlinecite{Kotani1996}], and it reproduces excellently the YTiO$_3$ and
LaTiO$_3$ spectra. The result for the Ti$_2$O$_9$ cluster is quite different from that of the TiO$_6$ cluster,
and matches very well the experimental Ti$_2$O$_3$ spectra including the high intensity satellite features
S1 and S2. These findings show that the Ti ions in YTiO$_3$ and LaTiO$_3$ are relatively isolated, while
in Ti$_2$O$_3$ they form electronically very strongly bonded pairs, fully consistent with the analysis for the
valence band spectrum discussed above.

\begin{figure*}
    \centering
    \includegraphics[width=14cm]{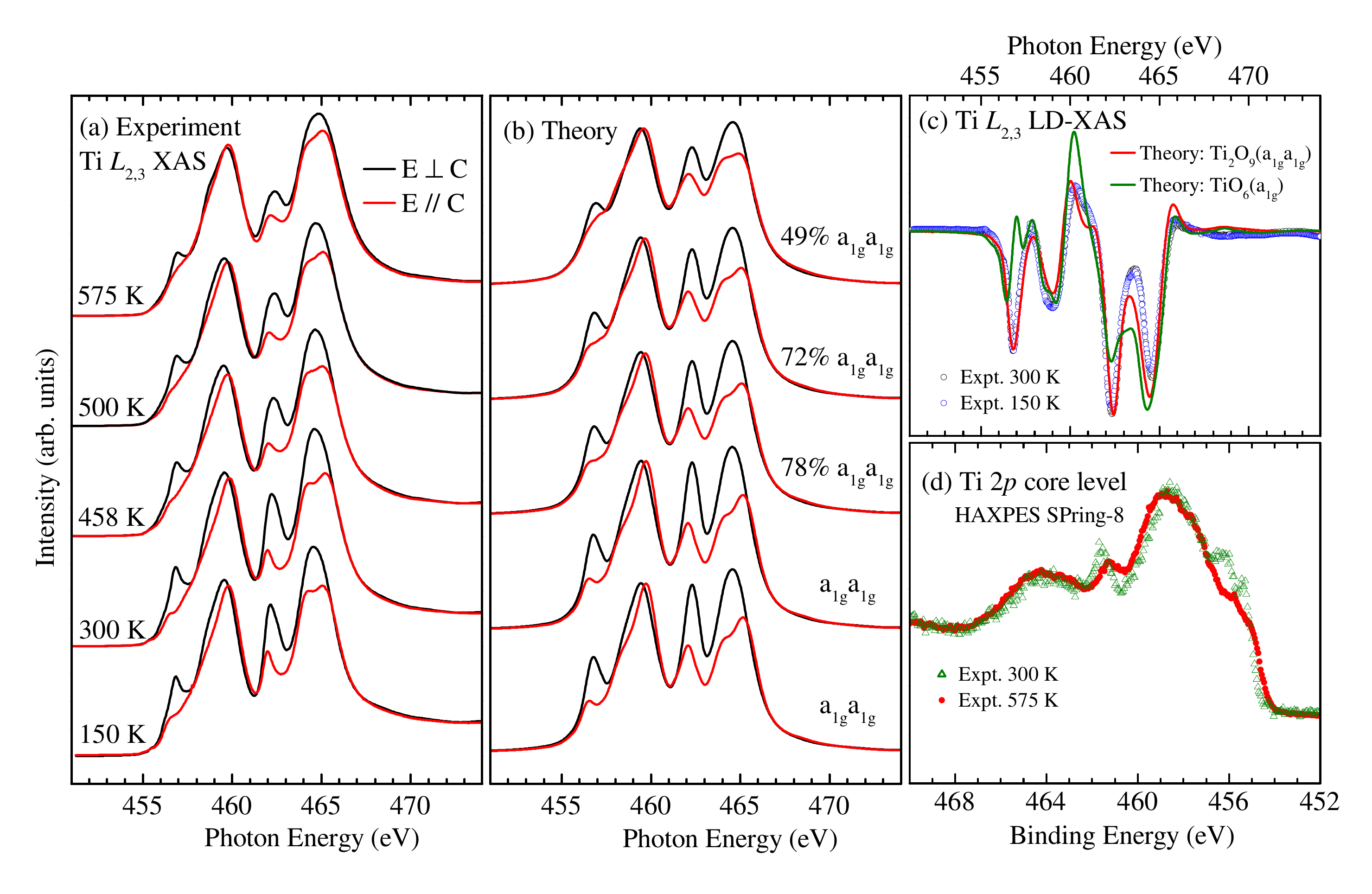}
    \caption{Panel (a): Experimental polarization-dependent Ti-$L_{2,3}$ XAS spectra of Ti$_2$O$_3$ taken
    at 150 K, 300 K, 458 K, 500 K and 575 K. Panel (b): Calculated polarization-dependent Ti-$L_{2,3}$ XAS
    spectra for the corresponding temperatures using a Ti$_2$O$_9$ cluster. Panel (c): Close up of the experimental
    linear dichroic (LD) spectrum in the low temperature phase (blue and black circles) and the simulation using
    a TiO$_6$ (green line: $a_{1g}$) and a Ti$_2$O$_9$ (red line: $a_{1g}$$a_{1g}$) cluster.
    Panel (d): temperature dependence of the Ti $2p$ core level spectrum.}
    \label{Fig:XAS_exp_cal}
\end{figure*}

Having established the electronic presence of the $c$-axis dimers, we need to determine or to verify that
the relevant orbitals which form the bond are the Ti $a_{1g}$. We need also to investigate how this orbital
occupation may evolve as a function of temperature across the insulator-metal transition.
The panel (a) of Fig.~\ref{Fig:XAS_exp_cal} shows the polarization-dependent Ti $L_{2,3}$ XAS spectra
of Ti$_2$O$_3$ taken at 150 K, 300 K, 458 K, 500 K and 575 K, i.e. from deep in the insulating low
temperature phase, across the gradual insulator-metal transition, and well into the metallic high temperature
phase. We can observe a strong polarization dependence indicative of a distinct orbital occupation of the
Ti $3d$ shell. We can also notice that the polarization dependence decreases across the transition.
In order to quantitatively extract the orbital occupation of the Ti $3d$ states from these Ti L$_{2,3}$ XAS spectra,
we simulated the spectra using the Ti$_2$O$_9$ cluster. The results are shown in the panel (b) of Fig.~\ref{Fig:XAS_exp_cal}.
We can clearly observe the excellent overall match between experiment and theory for all temperatures.

Focusing first on the low temperature phase, we find that the 150 K and 300 K spectra can be very well
described by a Ti-Ti $c$-axis dimer in an essentially pure $a_{1g}$$a_{1g}$ singlet ground state. Also a close-up
look at the dichroic spectrum, i.e. the difference between the spectrum taken with $\mathbf{E}$ $\|$ c
and the spectrum with $\mathbf{E}$ $\bot$ c spectra, where $\mathbf{E}$ denotes the electric field vector of
the incoming light, shows that the $a_{1g}$$a_{1g}$ ground state reproduces the experiment to a great detail,
see panel (c)  of Fig.~\ref{Fig:XAS_exp_cal} (experiment: blue and black dots; simulation: red line).
By contrast, a single-site TiO$_6$ cluster with a $a_{1g}$ initial state produces a significantly poorer fit
(green line). The low temperature XAS spectra
thus not only confirm fully the findings from the photoemission experiments shown above about the strong intra-dimer
electronic bond but also that this bond is formed by the $a_{1g}$$a_{1g}$ singlet.

With this finding we in fact restore the presumptions of the early model by Goodenough and van-Zandt \emph{et al.}
\cite{Goodenough1960,Zandt1968} for the insulating state, namely that the ground state is given by the dimer
in the $a_{1g}$$a_{1g}$ singlet. This model has been rejected for decades by band structure calculations
\cite{Zeiger-PRB-1975,Mattheiss1996,Eyert2005} on the basis that these calculations found a heavily mixed
orbital occupation. We also validate completely the starting point of the Mott-Hubbard model by Tanaka \cite{Tanaka2004},
thereby correcting the numbers found in an earlier polarization dependent Ti L$_{2,3}$ XAS experiment \cite{Hitoshi2006}:
by extending our experiment to lower temperatures, we ensure that our low temperature spectrum is taken from deep
inside the insulating phase, and by including the O 2p ligands in our analysis, we were able to obtain a better match
between the simulation and experiment as shown in panel (c) of Fig.~\ref{Fig:XAS_exp_cal}.

Concerning the temperature evolution, the decrease in the polarization dependence of the XAS spectra across
the gradual insulator-metal transition can be ascribed to a re-population of the Ti $3d$ orbitals \cite{Hitoshi2006}.
Our simulations find that the occupation of the $a_{1g}$$a_{1g}$ singlet state is reduced to 78\%, 72\%, and
49\% for $T$ = 458 K, 500 K and 575 K, respectively. The $e_g^{\pi}$ orbitals of the Ti $t_{2g}$ shell get more
and more occupied. The Ti ion thus becomes electronically less anisotropic, thereby also weakening and eventually,
breaking-up the electronic bond of the $c$-axis dimer. This is mirrored by the strong changes in the Ti $2p$ core-level
spectrum and in particular in the reduction of the satellites S1 and S2 intensities, see panel (d) of Fig.~\ref{Fig:XAS_exp_cal}.
Also the lengthening of the $c$-axis dimer bond distance \cite{Hwang2016} across the gradual insulator-metal transition
can be viewed as a weakening of the bond.

\begin{figure}
    \centering
    \includegraphics[width=8.5cm]{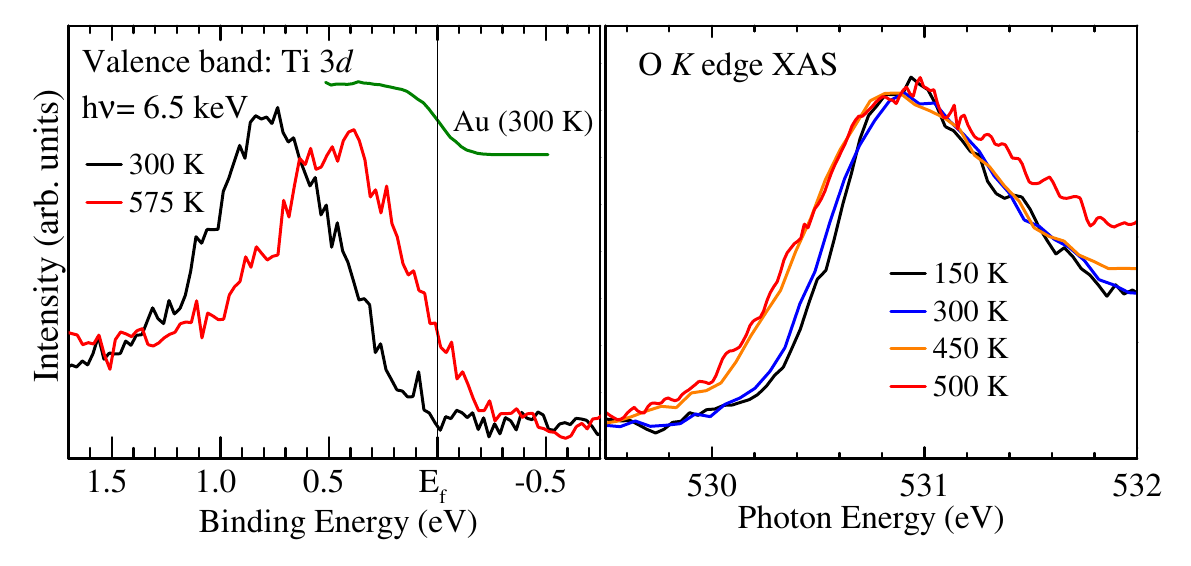}
    \caption{Left panel: Close-up of the temperature dependence of the valence band spectra of
     Ti$_2$O$_3$ taken with  $h\nu\simeq 6.5$ keV (HAXPES SPring-8) together with the Au Fermi
     cut-off as $E_F$ reference. Right panel: Close-up of the temperature dependence of the O $K$
     edge XAS spectra of Ti$_2$O$_3$.}
    \label{Fig4}
\end{figure}

The partial break-up of the dimer and the orbital reconstruction with temperature have consequences for the states
closest to the chemical potential. In the left panel of Fig.~\ref{Fig4} we display a close-up of the valence band
collected with the bulk-sensitive HAXPES method and in the right panel we show the threshold region of the O K
edge XAS as an indicator for the unoccupied states. We can clearly observe the gradual closing of the band gap.
Perhaps more striking is the fact that spectral weight is transferred over an energy range of 0.3 eV on both sides
of the chemical potential. Similar effects can also be seen in the optical conductivity \cite{Uchida2008}.
One may try to explain these changes in terms of large shifts in the energies of the relevant orbitals
\cite{Zandt1968,Uchida2008}, or perhaps also in terms of large changes in the strength of the intra-dimer
hopping integrals, but the density of states from band structure calculations
\cite{Zeiger-PRB-1975,Mattheiss1996,Iori-PRB-2012,Singh_JAlloysCompounds-2016} show only
modest changes with temperature.

We therefore have to take strong electron correlation effects explicitly into account in the explanation,
so that small changes in the one-electron band width and in the strength of the effective Coulomb
interaction $U_{eff}$ can lead to large changes in the electronic structure, e.g. a MIT with
a large transfer of spectral weight. The key issue for Ti$_2$O$_3$ is the orbital reoccupation
away from a pure $a_{1g}$$a_{1g}$ dimer singlet state at low temperatures.
Those dimers are electronically isolated from each other in the solid: the hopping in the $a$-$b$ plane
for electrons in the $a_{1g}$ orbital (oriented along the $c$-axis) is small ($\leq$ 1 eV
\cite{Eyert2005,Singh_JAlloysCompounds-2016}) compared to the Coulomb energy $U_{eff}$=$U$,
where $U$ ($\approx$ 3 eV, see above) is the energy repulsion between two electrons doubly
occupying the same $a_{1g}$ orbital of one particular Ti site after such a hopping process in the plane.

If on the other hand, the electrons are also allowed to occupy the $e_g^{\pi}$ orbitals, then the hopping in
the $a$-$b$ plane will be greatly enhanced, simply because the $e_g^{\pi}$ orbitals are much more directed
in this plane. Moreover, after such a hopping the doubly occupied state can be a triplet (high spin)
$a_{1g}$$e_g^{\pi}$.  The effective Coulomb energy  $U_{eff}$ will then be given by $U'$-$J_H$,
where $J_H$ denotes the gain in Hund's rule exchange energy for pairs of spin-parallel electrons, and
where $U'$ is the energy repulsion between electrons in different orbitals, to be distinguished from $U$
which is for electrons in the same orbital. $U'$ is smaller than $U$ by about 2$J_H$, so that $U_{eff}$ for the
$a_{1g}$$e_g^{\pi}$ situation is smaller by an amount of 3$J_H$ than the $U_{eff}$ for the $a_{1g}$$a_{1g}$.
Considering that $J_H$ is typically 0.7 eV \cite{Tanaka1994}, $U_{eff}$ for the $a_{1g}$$e_g^{\pi}$ can be
2.1 eV smaller than that for the $a_{1g}$$a_{1g}$. So for the $a_{1g}$$e_g^{\pi}$ situation the $e_g^{\pi}$ band
width in the $a$-$b$ plane ($\approx$ 1.5 eV \cite{Eyert2005,Singh_JAlloysCompounds-2016}) can overcome
$U_{eff}$ ($\approx$ 0.9 eV) as to stabilize a metallic state.

To justify that converting a dimer in the singlet $a_{1g}$$a_{1g}$ situation into a dimer with the triplet
$a_{1g}$$e_g^{\pi}$ requires only a modest amount of energy, we need to estimate the dimer's total energy
in each situation.
For the $a_{1g}$$a_{1g}$, the hopping between the state with one electron on each Ti and the
state with one of the Ti doubly occupied and the other empty is given by $\sqrt{2}t_{a1g}$ (there are two
ways to make the latter state, thus the $\sqrt{2}$ factor). The $U_{eff}$ is given by $U$, see above.
For the $a_{1g}$$e_g^{\pi}$, the hopping between the state with one electron on each Ti and the
state with one of the Ti doubly occupied and the other empty is given by $t_{a1g}$ (here we neglect
the hopping of the $e_g^{\pi}$ along the $c$-axis completely). The $U_{eff}$ is $U'$-$J_H$, see above.
Thus, whereas for the $a_{1g}$$a_{1g}$ both the hopping and $U_{eff}$ are large, for the
$a_{1g}$$e_g^{\pi}$ both the hopping and $U_{eff}$ are small. One can therefore argue that
together with lattice effects one can find physically reasonable parameters as to keep the total energy
difference between the two situations within 0.1 eV. These findings provide strong experimental support
for the theoretical model proposed by Tanaka \cite{Tanaka2004} to explain the MIT in Ti$_2$O$_3$.

To summarize, using a combination of photoelectron and polarized x-ray absorption spectroscopy
we were able to establish that the low temperature phase of Ti$_2$O$_3$ can be viewed
as a collection of isolated $c$-axis Ti-Ti dimers in the singlet $a_{1g}$$a_{1g}$ configuration.
Upon heating and crossing the gradual insulator-metal transition, the dimers start to break-up
with a reoccupation of the orbitals. The availability of the $e_g^{\pi}$ channel increases the
hopping within the $a$-$b$ plane. The smaller effective Coulomb interaction for the triplet
$a_{1g}$$e_g^{\pi}$ configuration facilitates the orbital reconstruction and the stabilization
of the metallic state at high temperatures.

We thank Lucie Hamdan for her skillful technical assistance.
The research in Cologne is supported by the Deutsche Forschungsgemeinschaft
through SFB608 and in Dresden through FOR1346.

$^{\ast}$ Current address: Department of Physics, University of Trento, Via Sommarive, 14 - 38123 Povo, Italy


\begin{thebibliography}{99}
\bibitem[{\citenamefont{CIpara}(2017)}]{CIpara}
\bibinfo{note}{TiO$_6$ cluster [eV]: $U_{dd}$ = 4.0, $U_{pd}$ = 5.5, $\triangle$ = 6.5,
10$Dq_{ionic}$=0.85, $\Delta_{trg}$ = -0.16 for an $e_{g}^{\pi}$ initial state and 0.16 for an $a_{1g}$ initial state,
hybridization $V_{eg}^{\sigma}$ = 3.5, $V_{eg}^{\pi}$ = 1.2, $V_{a_{1g}}$ = 0.9.
Ti$_2$O$_9$ cluster [eV]: The hopping integral between two Ti atoms is simplified as a $V_{dd\sigma}$ = 0.9,
which is a half of the energy difference between the 3$d$ main peak (M)
and the satellite (S) of the valence-band spectrum according to the $H_2$ Hubbard model.
Otherwise, all parameters are the same as those used for the TiO$_6$ cluster.
For calculating the 2$p$ core-level photoemission spectrum,
due to the presence of the core hole, $U_{pd}$-$U_{dd}$ is 5.5 eV for the Ti$_2$O$_9$ cluster.}
\end{thebibliography}
\end{document}